\documentstyle[12pt,aaspp4]{article}

\begin{document}

\title{Atomic Carbon Emission from Individual Molecular Clouds in M33}
\author{Christine D. Wilson\altaffilmark{1}}
\bigskip
\centerline{\it Submitted: June 6, 1997; revised July 2, 1997}  

\altaffiltext{1}{Department of Physics and Astronomy, McMaster University,
Hamilton, Ontario L8S 4M1 Canada }

\begin{abstract}

We present observations of the 492 GHz [CI] emission for four individual
giant molecular clouds in the Local Group galaxy M33 obtained
with the James Clerk Maxwell Telescope. The average
[CI] to CO J=1-0 integrated intensity ratio of $0.10\pm 0.03$
is similar to what is observed in Galactic molecular clouds but
 smaller than what is seen in starburst galaxies. Similarly,
the column density ratio $N(C)/N(CO)$ is similar to that observed
in the Orion Bar, but smaller than values obtained for starburst galaxies.
The [CI] line is found to be a more important coolant than the
lowest three rotational transitions of CO for all the clouds in the sample.
The [CI] luminosity does not appear to be enhanced significantly 
in two low-metallicity clouds, which may be due to 
the unusual ionization environment of the
clouds.

\end{abstract}


\section{Introduction}

The amount and distribution of atomic and ionized carbon are
important for understanding the physical and
chemical structure of molecular clouds,
as well as the effect of star formation in changing that structure.
Since the energy required to ionize carbon (11.3 eV) is close to that
required to photo-dissociate CO (11.09 eV), [CI] emission should exist
over a fairly narrow range of column densities, sandwiched between
regions where the carbon is predominantly ionized and regions where
it is bound up in CO (\markcite{t85}Tielens \& Hollenbach
1985). However, 
early results for the edge-on photo-dissociation region (PDR) in M17 
revealed that the [CI] emission is quite extended 
(\markcite{k85}Keene et al. 1985), while 
a more recent $25\times 25^\prime$ map  of S140 has confirmed
that the atomic carbon emission can be very extended indeed 
(\markcite{p94}Plume, Jaffe, \& Keene 1994). 
One interpretation for the large extent
of the [CI] emission is that the internal structure of molecular clouds
is clumpy, which allows
ultraviolet radiation to reach deep into
the cloud.
High-resolution maps in [CI] of a small area near the PDR
in M17  demonstrate that the [CI] emission
occurs on the edges of high-density clumps 
(\markcite{wp91}White \& Padman 1991) 
and provide direct confirmation that the clumpy structure of
the cloud is important to understanding the origin of the [CI] emission. 

The large extent of the [CI] emission in molecular clouds raises
the question of whether a PDR powered by
nearby massive star formation is necessary to produce significant
[CI] emission in a molecular cloud.
Because the [CI] emission is relatively insensitive
to the intensity of the ultraviolet radiation field
(\markcite{h91}Hollenbach, Tielens, \& Takahashi 1991), a substantial
amount of atomic carbon could exist  even in molecular clouds heated by the
 interstellar radiation field. Indeed, relatively large column densities
of atomic carbon are observed towards high latitude clouds
\markcite{svd94}(Stark \& van Dishoeck 1994).
Thus, rather than being confined to a small region near the PDR,
atomic carbon could be present on the entire
surface of the cloud, or even throughout the cloud depending on the 
self-shielding of the clumps and the amount of attenuation provided 
by the interclump medium. 
In addition, chemical processes involving
H$^+$ may be able to produce significant amounts of atomic
carbon throughout the cloud (\markcite{p92}Pineau des For\^ets, Roueff,
\& Flower 1992).

Observations of the total [CI] emission from individual  clouds
are the best way to determine just how widespread atomic carbon is
in molecular clouds. Unfortunately, 
large-area maps of [CI] emission are very time consuming to obtain with the
small beams of most submillimeter telescopes 
(see \markcite{p94}Plume et al. 1994 for
an interesting counter-example). However, the beams of the largest
submillimeter telescopes are sufficiently small that they can
isolate individual molecular clouds in galaxies in the Local
Group. Indeed, [CI] emission has been detected recently in
the LMC using a beam subtending 34 pc
(\markcite{sb97}Stark et al. 1997).
This paper presents the results of [CI] observations obtained
for four individual molecular clouds in the Local Group spiral
galaxy M33. 

\section{Observations and Data Reduction}

Four giant molecular clouds in M33 were observed with
the James Clerk Maxwell Telescope (JCMT) 
in the $^3P_1 - ^3P_0$ fine structure line of atomic carbon on 1995 October
29. The half-power beamwidth
of the JCMT at 492 GHz is 11\arcsec~ or 45 pc at
the distance of M33 (0.84 Mpc, \markcite{f91}Freedman, Wilson, \& Madore 1991).
The clouds were
selected from the interferometric samples of \markcite{ws90}Wilson \&
Scoville (1990, 1992). The observations were obtained
by position-switching with an offset of $<25^\prime$ to a point outside
the HI disk of the galaxy. 
The total integration time per cloud was 50 to 70 minutes. 
The data were binned to a resolution of 2 MHz (1.22 km s$^{-1}$) and
first order baselines (fifth for MC 20) were removed. 
All data were converted to the main beam
temperature scale using the   main beam efficiency at
 492 GHz published in the JCMT User's Guide ($\eta_{MB} = 0.52$). 
The absolute calibration  was checked by  observing the source W75N
which had a peak temperature of 12.1 K and an integrated
 intensity of 58.1 K km s$^{-1}$. This peak temperature is
$\sim$20\% smaller than the peak of 14 K found in standard JCMT
reference spectra.
The spectra  are shown in Figure~\ref{fig-1} and 
the measured [CI] properties are given in Table~\ref{tbl-1}. 

The integrated intensity for each cloud 
was measured by integrating over the full width of
the [CI] line, which for all clouds except MC 20 was somewhat smaller
than the width of the CO J=2-1 line observed with a 22$^{\prime\prime}$ beam
at the JCMT (\markcite{w97}Wilson, Walker, \& Thornley 1997).
Assuming the [CI] emission
is optically thin,  the atomic carbon column density is given by
$$N(C) = 2\times 10^{15} (e^{23.6/T_{ex}} + 3 + 5 e^{-38/T_{ex}})
\int T_{MB} dv \hskip6pt \rm{cm^{-2}}$$
(\markcite{ph81}Phillips \& Huggins 1981), where $T_{ex}$ is the
excitation temperature of the [CI] line and $\int T_{MB} dv$ is the
integrated intensity of the [CI] line. The minimum
column density is obtained for $T_{ex}$=24 K, while $N(C)$ is a factor
of two larger if $T_{ex}$ = 10 K and only 15\% larger if
$T_{ex}$=100 K. For the excitation temperature
we have adopted the kinetic temperature derived from a
large velocity gradient (LVG) analysis of
the low-J CO lines (\markcite{w97}Wilson et al. 1997). 
For two clouds (MC 20 and NGC 604-2), the LVG analysis gives 
 a range of kinetic temperature;
we have adopted the minimum temperature, which
minimizes both the atomic carbon and CO column
densities. Finally, since the sizes
of the clouds are known from the CO interferometer surveys, we have also
corrected $N(C)$  by the beam filling
factor, $f = (\overline{D}/
11^{\prime\prime})^2$, where $\overline{D}$ is the average full-width
half-maximum diameter of the cloud (\markcite{ws90}Wilson \& Scoville 1990,
1992).

To calculate the [CI] to CO integrated intensity ratio, $I_{[CI]}/I_{CO}$,
we used the published interferometric measurements of the CO J=1-0 transition 
(\markcite{ws90}Wilson \& Scoville 1990, 1992). The data were converted
from Jy to K using the conversion factor 
appropriate for each interferometric
beam ($\sim 0.6$ Jy/K).
In addition, since the beam of the interferometer
data is typically $8^{\prime\prime}\times 7 ^{\prime\prime}$, smaller
than the 11$^{\prime\prime}$ beam of the JCMT [CI] data, 
the CO data were scaled by the relative beam areas to obtain
temperatures appropriate for
an 11$^{\prime\prime}$ beam. 
The H$_2$ masses  were calculated from the CO
integrated intensities using  the 
calibration of \markcite{w95}Wilson (1995) to obtain a 
CO-to-H$_2$ conversion factor appropriate for the metallicity
of each cloud.

\section{Atomic Carbon Content of the Molecular Clouds in M33}

The average [CI] to CO integrated intensity ratio for this
sample of four clouds in M33 is $0.10\pm 0.03$. 
This line ratio is 
similar to that observed in Galactic molecular clouds, where
the $^{13}$CO J=1-0 and [CI] line intensities are comparable
(\markcite{t95}Tauber et al. 1995).
This  line ratio is lower than most values
observed in other galaxies,
which are typically in the range of 0.1-0.3 (Table~\ref{tbl-2}).
However, most measurements of $I_{[CI]}/I_{CO}$
have been of starburst galaxies, where the physical conditions are likely to be
quite different. The observations of M33 and the Milky Way 
suggest that $I_{[CI]}/I_{CO}$
is likely to be somewhat smaller in normal galaxies than in starburst galaxies.

The [CI] to CO integrated intensity ratio for the
 clouds in M33 ranges from
0.04 to 0.18 (Table~\ref{tbl-1}). 
For example,  the [CI] emission is four times
stronger in MC 20 than in MC 32 despite their
very similar masses and metallicities. 
Thus, we are seeing a real and significant
variation in the [CI] luminosity per unit mass  from one cloud to another
in M33. 
Recent observations of the LMC reveal similar line ratio 
variations  (\markcite{sb97}Stark et al. 1997). 
The higher [CI] luminosity of MC 20
is likely due to the presence of a nearby HII region
(\markcite{ws91}Wilson \& Scoville 1991), 
which probably produces a face-on PDR
 on the surface of the cloud. The two clouds near NGC 604 show
a similar trend, with  NGC 604-2, which is closest to the giant
HII region, having a larger [CI] luminosity 
than NGC 604-4.

The column density ratio $N(C)/N(CO)$ for three of the clouds 
is comparable to that observed
in the Orion Bar, 
but somewhat smaller than that observed in S140 and in
several starburst galaxies (Table~\ref{tbl-2}). 
The column density ratio is substantially lower for NGC 604-2, which
is subject to the most intense radiation field.
Unlike 30 Doradus in the LMC, where the [CI] emission is genuinely weak, 
the low $N(C)/N(CO)$ ratio in NGC 604-2 is due to a  large  CO column density
rather than a lack of [CI] emission.
This large value of $N(CO)$  arises from
the high kinetic temperature derived for NGC 604-2 (\markcite{w97}Wilson
et al. 1997). Given the unusual location of NGC 604-2 near the ionizing
star cluster of a giant HII region, it is quite likely that there
is more than one kinetic temperature component in this cloud. If
 the strong CO J=3-2 emission in this cloud originates in a small amount of
hot gas
while the lower transitions arise in a cooler component
(i.e. \markcite{w97}Wilson et al. 1997), the CO column density obtained
from the LVG analysis could be significantly overestimated.

There are a few possible sources of systematic error in our column
density analysis.  
First, although  we have assumed  the same beam-filling factors for 
the CO and [CI] emission,
the [CI] emission could be confined to a small portion of the cloud,
in which case $N(C)$ would be underestimated. This effect is
likely to be most important
 for the two clouds in NGC 604, which are subject
to edge-on heating by the giant HII region.
 Second, by
 adopting the lowest kinetic temperature
solution for MC 20 and NGC 604-2,
both $N(C)$ and $N(CO)$ may be underestimated.
Since $N(CO)$  is much more sensitive to temperature than $N(C)$,
 the net result would be an overestimate of $N(C)/N(CO)$.
Third, recent observations of the $^3P_2 - ^3P_1$ 809
GHz line of [CI] in M82 suggest that the optical depth in both [CI] 
lines may be close to unity  (\markcite{297}Stutzki et al.
1997). If the optical depths toward the M33 clouds are significant,
$N(C)$  would again be underestimated.
However, these effects are unlikely to be large enough to
bring the M33 line ratios into agreement with the starburst line ratios. 

Since MC 32 is located $\sim 120$ pc from the nearest HII region,
its atomic carbon  is likely
produced either by ionization
by the  interstellar radiation field 
(\markcite{h91}Hollenbach et al. 1991) or perhaps
by low density equilibrium chemistry (\markcite{p92}Pineau des For\^ets et 
al. 1992). The atomic carbon column density observed in MC 32 is $\sim 3$
times smaller than the column density predicted by models
of low-intensity PDRs ($N(C) \sim
10^{17}$ cm$^{-2}$ for $G/G_o=1$ and $n = 10^4$ cm$^{-3}$,
\markcite{h91}Hollenbach et al. 1991). However, these models assume
a solar gas-phase carbon abundance i.e. there is no depletion of carbon
onto dust grains. Models of translucent clouds ($1 < A_V < 5$)
assuming some depletion onto dust grains give
atomic carbon column densities similar to what is observed for
MC 32 and NGC 604-4 (\markcite{vdb88}van Dishoeck \& Black 1988,
\markcite{s96}Spaans 1996). The atomic carbon column densities
for MC 20 and NGC 604-2 agree quite well with dense PDR models with
moderate gas-phase carbon depletion
($N(C) \sim 1-6 \times 10^{17}$ cm$^{-2}$,
\markcite{vdb88}van Dishoeck \& Black (1988), but are again
somewhat smaller  than undepleted models
($N(C) \sim 4 \times 10^{17} - 2 \times 10^{18}$ cm$^{-2}$,
\markcite{th85}Tielens \& Hollenbach 1985). 

We have used our [CI] data to
estimate the relative importance of the [CI] and CO lines to the cooling
of molecular clouds.  Since the CO J=2-1 and J=3-2 measurements 
of these clouds were
made with larger beams (\markcite{w97}Wilson et al. 1997), we have chosen to 
scale the CO J=1-0 intensity in Table~\ref{tbl-1} 
 by the average CO J=2-1/J=1-0 
line ratio measured in M33 (0.67, \markcite{tw94}Thornley \& Wilson
1994) and by the CO J=3-2/J=2-1 line ratios appropriate for the individual
clouds (\markcite{w97}Wilson et al. 1997). With  all the
data now referred to the same beam, the integrated intensity ratios
can then be scaled by $(\nu_1/\nu_2)^3$ 
to convert from K km s$^{-1}$ to erg s$^{-1}$
cm$^{-2}$ sr$^{-1}$.
For all four clouds, the cooling in the [CI]
line exceeds that of the CO lines, by a factor of
1.5 for MC 32, $\sim 3$ for the NGC 604 clouds, and 7 for MC 20. 
The variations in the relative importance of the [CI] cooling are
primarily due to variations in the intensity of the [CI] line.
As expected,  [CI] cooling is most important  for MC 20,  the cloud
with a face-on PDR. 
However, the case of MC 32 shows that the 492 GHz line of [CI] is an important
coolant even for quiescent clouds without bright PDRs.

Since  atomic carbon may be more closely
associated with  atomic than  molecular hydrogen,
it is important to consider the  atomic hydrogen content of
the clouds.
 For MC~20,
the total mass of atomic hydrogen 
 is $1.5\times 10^4$ M$_\odot$, or about 5\% of the H$_2$ mass,
and is offset from the peak of the molecular gas by $\sim 70$ pc
(\markcite{ws91}Wilson \& Scoville 1991). 
Since the atomic carbon measurement was made towards the CO peak,
 it seems
very likely that, for MC 20, substantial atomic carbon is located in regions
that contain predominantly molecular rather than atomic hydrogen,
as is predicted by the
dense PDR models of \markcite{sd95}Sternberg \& Dalgarno (1995).
This conclusion is similar to recent results  for  ionized
carbon in  the LMC (\markcite{p95}Poglitsch et al. 1995), which
appears to be associated with molecular rather than atomic hydrogen.

The column density ratio $N(C)/N(H_2)$
is smaller by a factor of 2-3 in clouds with an oxygen abundance
reduced by 2.5 (Table~\ref{tbl-1}; 
compare MC 20 with NGC 604-2 and MC 32 with NGC 604-4).
This good correlation of $N(C)/N(H_2)$ with oxygen abundance
 suggests that the [CI] emission is at least marginally
optically thin and we are seeing the effect of the changing carbon
abundance in the clouds. In this  case, 
 the trend of carbon with oxygen in M33 would be more 
consistent with Galactic measurements, which show a constant C/O ratio
as the oxygen abundance decreases, than with measurements of
dwarf galaxies, where
the carbon abundance falls more quickly than the oxygen abundance
(\markcite{g95}Garnett et al. 1995). 
The low metallicity 
 of the two NGC 604 clouds could be responsible for
some  of the observed variation in the $I_{[CI]}/I_{CO}$ ratios
in M33. Lowering the metallicity by a factor of 2.5 reduces the CO
line strengths by a factor of $\sim 1.9$ (\markcite{w95}Wilson
1995). Assuming the [CI] intensity is linearly proportional to
the metallicity, the net result would be  to reduce $I_{[CI]}/I_{CO}$ for
the NGC 604 clouds by a factor of 1.3 compared to MC 20. 

It is often expected that
the size of the PDR
will be larger in low metallicity regions, due to 
reduced self-shielding and shielding by dust (i.e. \markcite{p95}Poglitsch 
et al. 1995). This effect is observed in the LMC, where the [CI] emission
appears enhanced at most of the positions observed to date
(\markcite{Sb97}Stark et al. 1997; Bania, private communication).
However, note that $I_{[CI]}/I_{CO}$ is what is observed to be
enhanced in the LMC; correcting for the lowered CO emission due to
the metallicity of the LMC would bring this ratio in good agreement
with the Milky Way value.
One interesting result of our study is that the two 
low-metallicity clouds in NGC 604
 have slightly smaller
[CI] luminosities than the more metal rich cloud MC 20.
The weak [CI] emission 
could be attributed to
 the intense ionization in the vicinity
of the giant HII region, which may tend to
convert the atomic carbon into ionized carbon 
(see the discussion of 30 Doradus in
\markcite{sb97}Stark et al. 1997). 
Other competing factors such as a drop in the absolute abundance of
carbon may be complicating the picture as well. 

\section{Conclusions}

We have detected [CI] emission from four individual giant
molecular clouds in the Local Group spiral galaxy M33. The
average [CI] to CO J=1-0 integrated intensity ratio is $0.10\pm 0.03$, 
similar to what is observed in Galactic clouds but somewhat lower than in
starburst galaxies. The [CI] emission is 
observed to be stronger for clouds associated with optical HII regions.
The column density ratio $N(C)/N(CO)$ for three of the clouds
is similar to that observed in the Orion Bar, but smaller than in
starburst galaxies. The cloud nearest the giant HII region NGC 604
has a value of $N(C)/N(CO)$ that is five times smaller than the other
clouds. This small column density ratio can be traced to a high
CO column density rather than a low atomic carbon column density.
The atomic carbon column densities for all clouds are in
reasonable agreement with predictions from
 photo-dissociation region models. The 
cooling by the 492 GHz [CI] line dominates the cooling in the lowest
three rotational transitions of CO, and
is important  even for clouds without active
star formation. Contrary to expectations,
the [CI] luminosity does not appear to be enhanced in two clouds
with lower metallicities. The unusual ionization environment of the
clouds or a drop in the absolute abundance of carbon may be
complicating the analysis.

\acknowledgments

I would like to thank Henry Matthews for performing the remote observing.
This research is supported through a grant
from the Natural Sciences and Engineering Research Council
of Canada. The JCMT 
is operated by the Royal Observatories on behalf of the Particle
Physics and Astronomy Research Council of the United Kingdom, the
Netherlands Organization for Scientific Research, and the National Research
Council of Canada.

\clearpage

\figcaption[fig1.ps]{$^3P_1 - ^3P_0$ [CI] 
spectra for individual molecular clouds
in M33 are shown as the heavy line.  The beam 
of the [CI] observations is 11$^{\prime\prime}$. The $^{12}$CO J=2-1 spectra
obtained with a 22$^{\prime\prime}$ beam are shown 
for comparison (thin line).
\label{fig-1}}

\clearpage

\begin{deluxetable}{lcccc}
\footnotesize
\tablecaption{Atomic Carbon Emission from Molecular Clouds in M33 
\label{tbl-1}}
\tablewidth{0pt}
\tablehead{
\colhead{Cloud} & \colhead{MC 20} & \colhead{MC 32}  
& \colhead{NGC 604-2}   & 
\colhead{NGC 604-4} \nl
} 
\startdata
$I_{[CI]}$$^a$ (K km s$^{-1}$) & $6.0\pm 0.7$ & $1.5\pm 0.4$ & $2.8\pm 0.4$ & 
$1.6\pm 0.3$ \nl
$I_{[CI]}/I_{CO}$$^b$ & $0.18\pm 0.05$ & $0.042\pm 0.014$ & $0.10\pm 0.03$ & 
$0.062\pm0.017$ \nl
$M_{H_2}$ ($10^5$ M$_\odot$) & 3.8 & 3.8 & 6.3 & 5.5 \nl
12+log(O/H)$^c$ & 8.89 & 8.89 & 8.48 & 8.48 \nl
H$\alpha$ flux$^d$  & $\ge 1$ & 0.05 & 4.8 & 0.95 \nl
$f^e$ & 0.78 & 1 & 0.56 & 0.93 \nl
$T_{ex}$$^f$ (K) & 20 & 10 & 100 & 10 \nl
$N(CO)$$^f$ (10$^{17}$ cm$^{-2}$) & 7 & 3 & 30 & 3 \nl
$N(C)$ (10$^{16}$ cm$^{-2}$) & 11 & 4.3 & 8.0 & 4.8 \nl
$N(C)/N({\rm H_2})$$^g$ ($10^{-6}$) & 11 & 6.5 & 3.5 & 3.4 \nl
$N(C)/N(CO)$ & 0.16 & 0.14 & 0.03 & 0.16 \nl
\enddata
\tablenotetext{a}{Uncertainties are the formal measurement uncertainties
only; the 20\% absolute calibration uncertainty is not included.}
\tablenotetext{b}{Ratio uses the CO J=1-0 line. The uncertainties
are calculated assuming a 20\% calibration uncertainty in each line.}
\tablenotetext{c}{\markcite{v88}Vilchez et al.
1988.}
\tablenotetext{d}
{Relative H$\alpha$ flux (erg s$^{-1}$ cm$^{-2}$) normalized to the
flux of MC 20.
Adopted H$\alpha$ luminosities and projected
distances from the molecular clouds 
are: for MC 20, $1.1 \times 10^{38}$ erg s$^{-1}$ and 25 pc;
for MC 32, $1.4 \times 10^{38}$ erg s$^{-1}$ (to MC 35 HII region) and 25 pc;
for NGC 604-2, $5.6\times 10^{39}$ erg s$^{-1}$ and 80 pc; 
for NGC 604-4,  $5.6\times 10^{39}$ erg s$^{-1}$ and 180 pc
(\markcite{ws91}Wilson \& Scoville 1991, 1992).}
\tablenotetext{e}
{Filling factor of the CO J=1-0 emission mapped interferometrically
(\markcite{w90}Wilson \& Scoville 1990, 1992) 
within the 11$^{\prime\prime}$ [CI] beam.}
\tablenotetext{f}
{Only the lowest possible excitation temperature and $N(CO)$
are given for MC 20 and NGC 604-2
(\markcite{w97}Wilson et al. 1997).}
\tablenotetext{g}{H$_2$ column densities are
area-averaged values calculated from the H$_2$ mass and the
area of the cloud.}
\end{deluxetable}

\begin{deluxetable}{lccl}
\footnotesize
\tablecaption{Comparison with Previous [CI] Observations
\label{tbl-2}}
\tablewidth{0pt}
\tablehead{
\colhead{Source} & \colhead{$I_{[CI]}/I_{CO}$$^a$} 
& \colhead{$N(C)/N(CO)$} & \colhead{References} \nl
} 
\startdata
M82 & 0.11 & 0.5 & Schilke et al. 1993, Stutzki et al. 1997 \nl
IC 342 & 0.17 & ... & B\"uttgenbach et al. 1992 \nl
NGC 253 & 0.30 & 0.2-0.3 & Israel et al. 1995 \nl
Cloverleaf & 0.18$^b$ & $\sim 1$ & Barvainis et al. 1997 \nl
Milky Way & 0.16$^c$ & ... & Wright et al. 1991 \nl
LMC & 0.1,0.26 & ... & Stark et al. 1997 \nl
M33 & 0.04-0.18 & 0.03-0.16 & this paper \nl
Orion & ... & 0.17 & Tauber et al. 1995 \nl
S140 & ... & 0.5 & Plume et al. 1994 \nl
\enddata
\tablenotetext{a}
{CO J=1-0 line. Integrated intensities are in 
units of K km s$^{-1}$.}
\tablenotetext{b}
{CO J=3-2 line.}
\tablenotetext{c}
{Calculated from the CO J=2-1 line assuming a
CO J=2-1/J=1-0 ratio of 0.7 (\markcite{s94}Sakamoto et al. 1994).}
\end{deluxetable}


\clearpage
 
\begin{references} 
\reference{b97}Barvainis, R., Maloney, P., Antonucci, R., \& Alloin, D.,
1997, ApJ, 484, 695

\reference{b92} B\"uttgenbach, T. H., Keene, J., Phillips, T. G., \& Walker,
C. K.,  1992, \apj, 397, L15

\reference{f91}Freedman,  W. L., Wilson,  C. D., \& 
Madore, B. F., 1991, \apj, 372, 455

\reference{g95} Garnett, D. R., Skillman, E. D., Dufour, R. J., Peimbert, M.,
Torres-Peimbert, S., Terlevich, R., Terlevich, E., \& Shields, G. A., 
  1995, \apj, 443, 64

\reference{h91} Hollenbach, D. J., Tielens, A. G. G. M., Takahashi, T., 
 1991, \apj, 377, 192


\reference{i95} Israel, F. P., White, G. J., \& Baas, F.,  1995, \aap, 295, 599

\reference{k85} Keene, J., Blake, G. A., Phillips, T. G., Huggins, P.
 J., \& Beichman, C. A. 1985, ApJ, 299, 967
 
\reference{ph81}Phillips, T. G., \& Huggins, P. J., 1981, \apj, 251, 533

\reference{p92} Pineau des For\^ets, G., Roueff, E., \& Flower, D. R., 1992, 
\mnras, 258, P45

\reference{p94} Plume, R., Jaffe, D. T., \& Keene, J.,  1994, \apj, 425, L49

\reference{p95}Poglitsch, A., Krabbe, A., Madden, S. C., Nikola, T.,
Geis, N., Johansson, L. E. B., Stacey, G. J., \& Sternberg, A., 1995,
\apj, 454, 293 

\reference{s93}  
Schilke, P., Carlstrom, J. E., Keene, J., \& Phillips, T. G.,  1993, \apj,
 417, L67

\reference{s94}Sakamoto, S., Hayashi, M., Hasegawa, T., Handa, T., \&
Oka, T., 1994, \apj, 425, 641

\reference{s96}Spaans, M., 1996, \aap, 307, 271

\reference{sb97} Stark, A. A., 
Bolatto, A. D., Chamberlin, R. A., Jackson, J. M., Lane,
A. P., \& Bania, T. M., 1997, \apj, 480, L59

\reference{svd94}Stark, R., \& van Dishoeck, E. F., 1994, \aap, 286, L43

\reference{sd95} Sternberg, A., \& Dalgarno, A., 1995, \apjs, 99, 565

\reference{s97} Stutzki, J., et al., 1997, \apj, 477, L33

\reference{t95} Tauber, J. A., Lis, D. C., Keene, J., Schilke, P., 
B\"uttgenbach,
T. H.,  1995, \aap, 297, 567

\reference{tw94}Thornley, M. D., \& Wilson,  C. D.
1994, \apj,  421, 458

\reference{t85}Tielens, A. G. G. M., \& Hollenbach, D., 1985, \apj, 291, 722

\reference{vdb88}van Dishoeck, E. F., \& Black, J. H., 1988, \apj, 334, 771

\reference{v88} Vilchez, J. M., Pagel, B. E. J., Diaz, A. I., Terlevich, E., 
\& Edmunds, M. G., 1988, \mnras, 235, 633
 
\reference{wp91} White, G. J., \& Padman, R. 1991, Nature, 354, 511

\reference{w95}Wilson, C. D.,  1995, \apj, 448, L97

\reference{ws90} Wilson, C. D., \& Scoville, N., 1990, \apj, 363, 435

\reference{ws91}Wilson, C. D.,  \& Scoville, N., 1991, \apj, 370, 184

\reference{ws92} Wilson, C. D., \& Scoville, N., 1992, \apj, 385, 512

\reference{w97} Wilson, C. D., Walker, C. E., \& Thornley, M. D., 1997, \apj,
483, 210

\reference{w91}Wright, E. L., et al., 1991, \apj, 381, 200


\end{references}
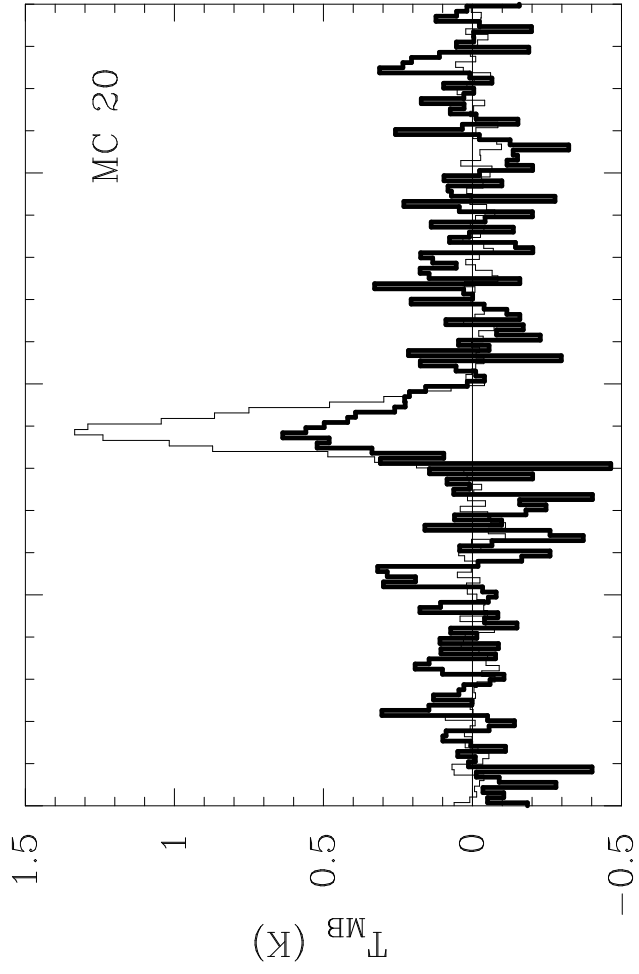
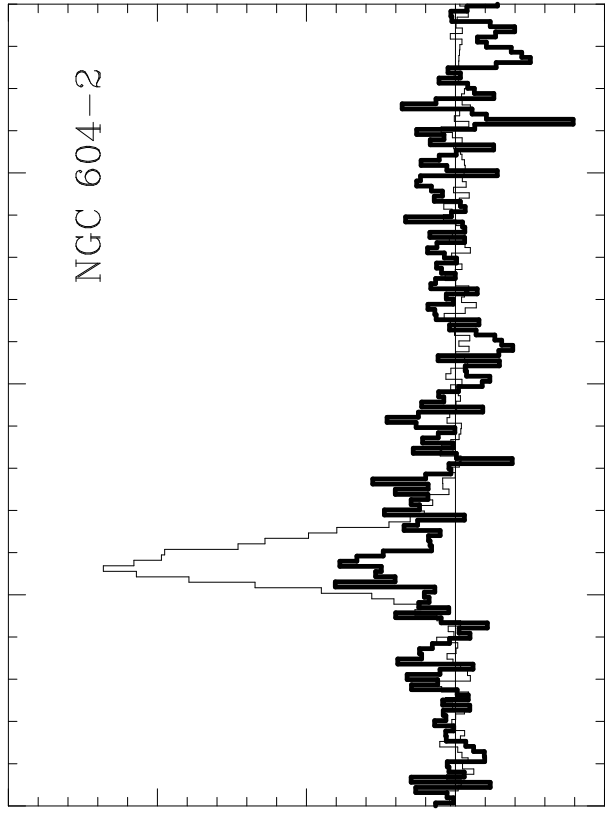
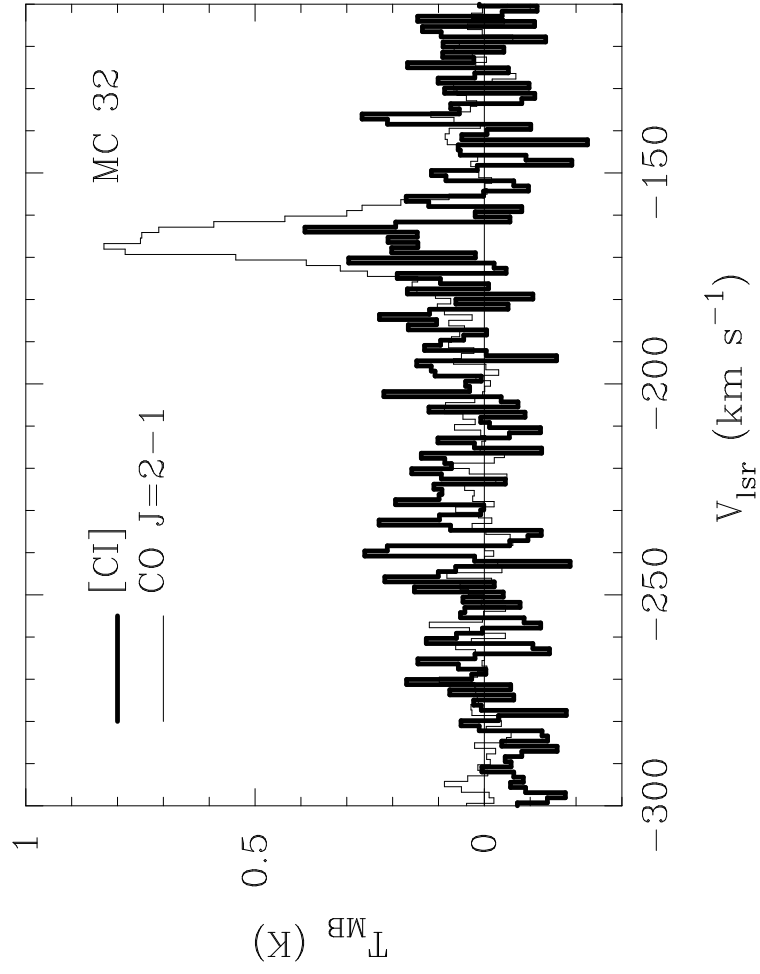
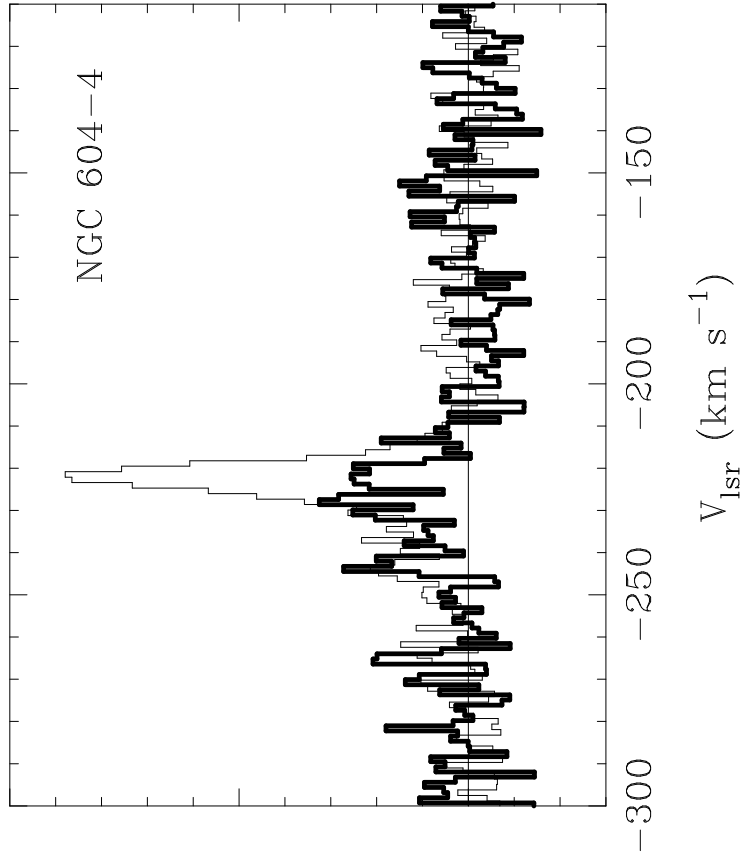
\end{document}